\documentclass[a4paper,11pt]{article}
\pdfoutput=1 

\usepackage{jheppub} 
\usepackage{subfigure}
\usepackage{xcolor}
\usepackage[T1]{fontenc} 
\usepackage{xspace}
\usepackage{multirow}
\usepackage{hyperref}
\usepackage{slashed}
\usepackage{amsmath}

\usepackage[utf8]{inputenc}

\def\GeV{\ifmmode {\mathrm{\ Ge\kern -0.1em V}}\else \textrm{Ge\kern -0.1em V}\fi}%
\def\GeV{\ifmmode {\mathrm{\ Ge\kern -0.1em V}}\else \textrm{Ge\kern -0.1em V}\fi}%

\title{\boldmath Scattering Amplitudes of Fermions on Monopoles}

\author{Valentin V. Khoze}

\affiliation{IPPP, Department of Physics, Durham University, Durham DH1 3LE, UK}

\emailAdd{valya.khoze@durham.ac.uk}

\abstract{
We consider scattering processes involving massless fermions and 't Hooft--Polyakov magnetic monopoles in a minimal SU(2) model and in the Grand Unified SU(5) 
theory. We construct expressions for on-shell amplitudes for these processes in the $J=0$ partial wave using the spinor helicity basis consisting of single-particle 
and pairwise helicities. These processes are unsuppressed and are relevant for the monopole catalysis of proton decay. The amplitudes for the minimal processes involving a single fermion scattering on a monopole in the initial state and half-fermion solitons in the final state
are presented for the first time and are used to obtain the amplitudes for processes involving more fermions in the initial state and integer fermion numbers in the final state. A number of such anomalous and non-anomalous processes, along with their amplitude expressions, are written down for the $SU(5)$ GUT model.}

\begin{document}
\preprint{IPPP/23/XX}

\maketitle
\flushbottom


\section{\label{Sec:Intro}Introduction}

Studies of scattering processes of light electrically charged fermions on magnetic monopoles, pioneered in the early eighties by 
Rubakov and Callan~\cite{Rubakov:1981rg,Callan:1982ah,Callan:1982au},
provide some remarkable insights into fundamental properties of quantum field theory.

It is well-known that no Lagrangian formulation exists which is both Lorentz-invariant and local, for a QFT with electric and magnetic degrees of freedom.
Zwanziger's formulation~\cite{Zwanziger:1970hk} has the advantage of featuring a {\it local} Lagrangian which depends on two gauge potentials and on an external
vector $n_\mu$ associated with the Dirac string.
The monopoles here are of Dirac type~\cite{Dirac:1931kp} given by external point-like U(1) magnetic sources 
with no non-singular core unlike the  finite-energy 't~Hooft--Polyakov monopole solutions~\cite{tHooft:1974kcl,Polyakov:1974ek} in a non-Abelian theory.
For certain applications Zwanziger's theory can be used successfully as a low-energy EFT description of the underlying fundamental non-Abelian theory,
but it fails to describe scattering processes of light fermions on 't Hooft--Polyakov monopoles even at low energies.

When light elementary fermions scatter on a 't Hooft-Polyakov monopole, the fermions in the $J=0$ partial wave can penetrate the monopole core even 
at asymptotically low energies and probe the short-distance non-Abelian dynamics of the underlying microscopic theory.\footnote{In general, for a fermion scattering on a monopole, the lowest partial wave in the partial wave expansion of the amplitude~\cite{Wu:1976ge,Kazama:1976fm}, is characterised by $J=j_0$ where $j_0=|q|-1/2$ and $q$ is the product of the fermion's electric and the monopole's magnetic charge over $2\pi$. For the SU(2) and GUT models considered in this paper we have $j_0=0$, {\it cf.}~Eq.~\eqref{eq:qdef}. For brevity we in this paper will refer to them as $J=0$ amplitudes.}
These $J=0$ amplitudes saturate unitarity and the cross-sections for the corresponding $s$-channel processes behave as
 $\sigma_s \propto 1/p_c^2$ for $p_c \gtrsim \Lambda_{QCD}$, where $p_c$ is the fermion 3-momentum in the CoM frame~\cite{Kazama:1976fm,Rubakov:1981rg,Callan:1982ah,Callan:1982au}. 
These results are to be contrasted with 
what one would have naively expected to hold in a perturbative QFT settings where  an $s$-channel process would have a $1/s$ pole with the corresponding cross-section for a head on collision being suppressed by $1/M_X^2 \sim R_M^2 $ at low momenta, where $R_M$ is the monopole core size and $M_X$ is the mass of vector bosons in the non-Abelian theory. 
It turns out instead that the fermion--monopole scattering in the spherical wave is in fact unsuppressed by the fundamental mass scale of the microscopic theory. 

In particular, for monopoles in a Grand Unified Theory (GUT) there are unsuppressed anomalous scattering processes of the type~\cite{Rubakov:1981rg,Callan:1982ah},
\begin{equation}
\label{eq:m-cat-p}
 u^1 \,+\,u^2 \,+\,  M \,\to\, \bar{d}^{3} \,+\, \bar{e} \,+\,  M,
\end{equation}  
which lead to the monopole catalysis of proton decay. This is the seminal Rubakov--Callan effect.
For GUT monopoles the natural scale of the $J=0$ scattering process is the scale of strong interactions, some 15 orders of magnitude below the relevant GUT scale that gave rise to 't Hooft-Polyakov monopoles in the first place. 

One can also consider a single fermion, for example a massless positron, scattering on a GUT monopole~\cite{Callan:1982au},
\begin{equation}
\label{eq:callan}
 \bar{e} \,+\,  M \,\to\, \frac{1}{2} \left(u^1\,u^2\,d^3\,\bar{e} \,\right) \,+\,  M.
\end{equation} 
Here the Callan bosonization formalism for $J=0$ scattering implies that particles in the final state carry half-integer fermion numbers, 
which in the first instance appears to be highly counter-intuitive as these cannot be described by the usual asymptotic non-interacting Fock states in perturbation theory.
We will study such processes in detail.

\medskip

The fact that there exists no acceptable Lagrangian formulation describing QFT interactions of electric particles and magnetic monopoles should by itself not be an  obstacle for studying their scattering amplitudes.
In fact, the lack of the underlying Lagrangian along with the apparent strong-coupling regime for the magnetic monopole coupling implied by the  
Dirac-Schwinger-Zwanziger (DSZ) quantization condition~\cite{Dirac:1931kp,Schwinger:1975ww,Zwanziger:1968rs},
makes an excellent case for developing an on-shell $S$ matrix formalism with no possible input from Feynman rules in a theory with electrically and magnetically charge particles.
This programme was carried out in a series of very interesting recent papers by Cs\'aki and collaborators~\cite{Csaki:2020inw,Csaki:2020yei,Csaki:2022tvb}.

At the foundation of this on-shell $S$ matrix formalism is the observation by Zwanziger 
 that asymptotic in- and out-states containing electrically charged particles and magnetic monopoles are not the tensor products of single particle states. Even at infinite separations between the electric charges and the monopole, there exists a non-vanishing angular momentum of the electromagnetic field~\cite{Thomson1904} and the electrically and magnetically charged states are pairwise entangled~\cite{Zwanziger:1972sx}. The  formalism developed in~\cite{Csaki:2020inw,Csaki:2020yei} provides a compelling practical realisation of this electric-magnetic entanglement by identifying the orbital momentum 
 carried by each pair of electrically and magnetically charged particles with a novel pairwise helicity variable associated with the pair.
 
 \medskip
 
Our goal in the this paper is to re-visit scattering processes of massless fermions on monopoles and to compute their amplitudes
in the regime where they are unsuppressed i.e. in the $J=0$ partial wave.
For the first time we will construct the amplitudes for the elementary processes of the type~\eqref{eq:callan} that involve a single fermion in the initial state and
fractionally charged fermions in the final state. We will then combine such processes to re-derive scattering amplitudes for the Rubakov--Callan
reactions with two fermions in the initial state~\eqref{eq:m-cat-p} thus validating the process~\eqref{eq:callan}.
Our results for the amplitudes of two fermions scattered on a monopole agree with the recent work~\cite{Csaki:2022qtz} but we disagree with their results for the processes 
where a single initial fermion is scattered on a monopole.

\medskip  
There is a number of intriguing and, at first sight, unexpected in QFT features that emerge from studying fermion--monopole scattering processes, which explains
the ongoing interest in these investigations:
\begin{enumerate}
\item{} There is no crossing symmetry. One can neither apply crossing  to individual particles in the Rubakov--Callan 
processes~\eqref{eq:m-cat-p}-\eqref{eq:callan}, nor would it be allowed by the multi-particle electric--magnetic entanglement;
\item{} Forward scattering amplitudes trivially vanish for many such processes, and
\item{} The optical theorem does not apply;

For example the complex conjugate amplitude for~\eqref{eq:m-cat-p} is the amplitude for
\begin{equation}
\label{eq:m-cat-p2}
\bar{d}^{3} \,+\, \bar{e} \,+\,  \overline{M} \,\to\, 
 u^1 \,+\,u^2 \,+\,  \overline{M},
\end{equation}  
which involves anti-monopoles rather than monopoles while the fermion states are the same.

\item{} As noted already, there is no decoupling of heavy mass-scales from the low-energy physics in fermion--monopole scattering. Low-energy fermions in the lowest partial wave penetrate freely non-Abelian monopole cores and result in unsuppressed scattering rates that cannot be obtained from low-energy U(1) EFT formulations.

\item{} There are fermion number violating anomalous, as well as fermion number preserving non-anomalous processes on monopoles that are both unsuppressed.

\item{} Production of fractional fermion numbers is possible for massless fermions scattered on monopoles thus restructuring the perturbative Fock space.

\end{enumerate}

\medskip

The paper is organised as follows.
In section~\ref{Sec:su2} we start with a minimal SU(2) theory that supports 't Hooft--Polyakov monopoles and 
provides simple settings and convenient general notation for studying all relevant to us aspects of fermion--monopole scattering.
Section~\ref{sec:2.1} presents a brief overview of fermion--monopole interactions and the special role played in the scattering by the $J=0$ partial wave.
The simplest model with $N_f=2$ massless fermion flavours and the scattering processes therein are discussed in section~\ref{sec:2.2}.
A more non-trivial case with $N_f=4$ flavours, which is also more relevant phenomenologically, is detailed in section~\ref{sec:2.3}.
In section~\ref{sec:2.4} we present our main results for the corresponding fermion--monopole scattering amplitudes using pairwise and single-particle helicity spinors.
Section~\ref{Sec:su5} summarises applications of our results to an SU(5) GUT theory with one generation of massless fermions and lists 
a number of scattering processes and their amplitudes. Section~\ref{Sec:concl} outlines our conclusions.


\section{Fermion--monopole scattering in the SU(2) Model}
\label{Sec:su2}

In this section we consider a basic model supporting 't Hooft-Polyakov monopoles: an SU(2) gauge theory with the Higgs field in the adjoint representation. 
We add $N_f$ flavours of Left-handed Weyl fermion doublets, which we take to be massless, 
\begin{equation}
\label{eq:ferm_ab}
\psi_L^{\,i}\,=\, \begin{pmatrix}
a_{+}^{\,i} \\
b_{-}^{\,i}
\end{pmatrix}_L\,, \quad i=1,\ldots N_f.
\end{equation}
Being elementary fields, the fermions are much lighter than the monopole, $m_{\rm ferm} \ll M_X/ \alpha$, where $M_X$ is the vector-boson mass\footnote{We use $X$ to denote massive vector bosons of the SU(2) model in analogy to heavy vector bosons in Grand Unification, which is a principal application for heavy monopole--fermion scattering.}
 and $\alpha=g^2/4\pi$ in the SU(2) theory. Hence we can set $m_{\rm ferm} =0$ is a good approximation.
 
We follow Rubakov's notation~\cite{Rubakov:1988aq} where Left-handed fermions $a_{+}$ and $b_{-}$ have respective U(1) electric charges $e =\pm 1/2$ in the units of the SU(2) gauge coupling $g$.
Their corresponding anti-particles, $\overline{(\psi_L^{\,i})} = \bar\psi_{\,\,R}^{\,i}$, transform as Right-handed spinors and are represented by the SU(2) doublets,
\begin{equation}
\label{eq:barferm_ab}
\bar\psi_{\,\,R}^{\,i}\,=\, \begin{pmatrix}
\bar{b}_{+}^{\,i}  \\
\bar{a}_{-}^{\,i}
\end{pmatrix}_R\,.
\end{equation}
The Witten's SU(2) anomaly argument requires that $N_f$ must be even and we will consider in turn the $N_f=2$ and $4$ cases, with the latter being particularly relevant to applications to Grand Unification in section~\ref{Sec:su5}.

The 't Hooft-Polyakov monopole~\cite{tHooft:1974kcl,Polyakov:1974ek} is a topological soliton and has an integer-valued topological charge $g_M \in {\cal Z}$. At distances much greater that the monopole core $r \gg R_M  \sim 1/M_X$, the 't Hooft-Polyakov solution in the unitary gauge coincides with the Dirac monopole configuration with $g_M$ being twice the magnetic charge of the Dirac monopole in units of $4\pi/g$.
Throughout the paper we consider the minimal 't Hooft-Polyakov monopole with the positive minimal value of the magnetic charge, $g_M= + 1$,
of the unbroken U(1).\footnote{In our conventions $g_M=1$ under the unbroken U(1) of the Georgi-Glashow SU(2) theory, and under the unbroken $U(1)_M$ subgroup of the Grand Unified SU(5) theory. For the latter model we note 
that the QED-electric charges of the fermions and the QED-magnetic charge of the monopole are opposite to those of the unbroken $U(1)_M$. These are the same conventions as the ones used by Rubakov and others~\cite{Rubakov:1981rg,Rubakov:1988aq,Preskill:1984gd}.}

It is easy to see that our monopole magnetic charge, $g_M=1$, times the electric charge of the $a_{+}$ or $b_{-}$ fermions satisfies,
\begin{equation}
\label{eq:qdef}
q\, \equiv\, e_{a_+/b_-} \cdot g_M =\, \pm \frac{1}{2} \,, \quad 
\end{equation}
in agreement with the standard Dirac quantization~\cite{Dirac:1931kp} condition, $q_{\rm el} \cdot q_{\rm mg} = 2 \pi n$, for the minimal value of the integer, $n=1$, after one recovers the units of gauge coupling $g$ and $4\pi/g$ for the electric and magnetic charges.

\subsection{Unsuppressed s-wave interactions inside the monopole core}
\label{sec:2.1}
\medskip

When massless fermions scatter on a static 't Hooft-Polyakov monopole, the fermions in the $J=0$ partial wave can penetrate the monopole core, while the higher $J>0$  harmonics experience a centrifugal barrier in their interaction potential with the monopole and scatter without being able to reach inside the monopole core. 
This implies that the fermion--monopole scattering in the spherical $J=0$ wave saturates unitarity and
 is unsuppressed by the monopole core mass scale. This observation is the first of the two key features of fermion--monopole scattering that are at heart of the 
 Rubakov--Callan  discovery of monopole catalysis of proton decay~\cite{Rubakov:1981rg,Callan:1982ah,Callan:1982au}.

The second key feature of the Rubakov--Callan approach is the observation that for Left-handed Weyl fermions in the $J=0$ wave only their $a_{+}$ components exist as incoming waves while their $b_{-}$ components give the outgoing states in the fermion--monopole scattering. For the Right-handed spinors, $\bar{a}_{-}$ are incoming and $\bar{b}_{+}$ are outgoing. These facts follows from truncating the theory to $J=0$ waves for each fermion, 
and analysing solutions of the Dirac equation for massless Weyl fermions in the 't Hooft--Polyakov monopole background. The truncation to $J=0$ partial waves
is justified since the higher partial waves cannot penetrate the monopole core and result in subdominant contributions to the scattering suppressed by powers of $M_X$
 (see e.g. comprehensive  reviews~\cite{Rubakov:1988aq,Callan:1983tm,Preskill:1984gd} for details of the derivation).

\medskip 

For future convenience, in Table~\ref{Tab:1} we summarise the charges~\eqref{eq:qdef} and properties of massless Weyl fermions~\eqref{eq:ferm_ab}-\eqref{eq:barferm_ab}
scattered on a 't Hooft--Polyakov monopole in $J=0$ partial wave.
 \begin{table}[ht]
\centering
\begin{tabular}{c  c c}
\hline\hline
$\psi_L$ & $\,\,q=eg_M$ & \,\,in/out  \\ [0.5ex] 
\hline
$a_{\,+L}$  & $1/2$ &{in} \\
$b_{\,-L}$  & $-1/2$ & {out} \\
[1ex]   
\hline\\
\hline\hline
$(\overline{\psi})_{R}$ & $\,\,q=eg_M$ & \,\,in/out \\ [0.5ex] 
\hline
 $\bar{b}_{\,+R}$ & $1/2$ &out \\
 $\bar{a}_{\,-R}$ & $-1/2$ &in \\
[1ex]   
\end{tabular}
\caption{s-wave (anti)-fermions scattering on a 't Hooft--Polyakov monopole of charge $g_M=1$.}
\label{Tab:1}
\end{table}

\subsection{Scattering in the $\mathbf SU(2)$ theory with $\mathbf N_f=2$ flavours}
\label{sec:2.2}
\medskip

In a scattering process in a gauge theory, electric charge must be conserved. For low-energy scattering processes we are interested in, that is when the energy carried by the initial state fermions is (much) lower than the monopole--dyon mass splitting, $\sim M_X \sim 1/R_M$, the electric charge cannot be deposited on the monopole core by turning it to a dyon.

Starting with a single fermion $a_{+L}^1$  in the initial state, the $J=0$ state amplitude on a monopole that satisfies the in/out selection rules of Table~\ref{Tab:1} and the electric charge conservation, allows for the process,
\begin{equation}
\label{eq:nf2-1f1}
 a_{+L}^1 \,+\, M \,\to\, \bar{b}_{\,+R}^{\,2} \,+\, M \,+\, (\bar{b}b)\,{\rm pairs} \,,
\end{equation}
in the $N+f=2$ model.
This is an anomalous process as the chirality is not conserved. In fact, by applying large gauge transformations to a static monopole configuration, one can construct a semiclassical monopole-state `theta-vacuum' as a linear combination of monopole states with different Chern-Simons numbers $n$. Instanton tunnelling effects that change $n\to n+1$ also change the number of R-handed minus L-handed fermions by one unit for each fermion flavour.
One can use a simple selection rule~\cite{Rubakov:1988aq} which states,
\begin{equation}
\label{eq:anom_n}
\Delta R^i - \Delta L^i = n \,,\quad {\rm for\,\,each\,\,flavour\,\, } i=1,\ldots,N_f \,,
\end{equation}
where $\Delta R^i$ and $\Delta L^i$ are the change in the Right- and Left-handed fermions between the final and the initial states for each flavour $i$, and $n$ a fixed constant for all $N_f$ flavours.
The process \eqref{eq:nf2-1f1} gives $\Delta R^1 - \Delta L^1 = 1 = \Delta R^2 - \Delta L^2$ and is allowed by the triangle anomaly. 

On the other hand, an alternative proposal for the scattering,  $a_{+L}^1 \,+\, M \,\to\, \bar{b}_{\,+R}^{\,1} \,+\, M$, is not allowed by the anomaly selection rule~\eqref{eq:anom_n} since in this case $n$ is not universal for different flavours:
$\Delta R^1 - \Delta L^1 = 2$ is not equal to $\Delta R^2 - \Delta L^2 = 0$. This process can also be ruled out by the flavour symmetry of the massless 2-flavour model at hand.
 
 \medskip
 
 The process~\eqref{eq:nf2-1f1} can also be used as an elementary building block for constructing multi-fermion--monopole scattering reactions in the $N_f=2$ model. The process with two fermions in the initial state incident on the monopole, 
\begin{equation}
\label{eq:nf2-2f2}
 a_{+L}^1 \,+\,a_{+L}^2 \,+\,  M \,\to\, \bar{b}_{\,+R}^{\,1} \,+\, \bar{b}_{\,+R}^{\,2} \,+\,  M \,+\, (\bar{b}b)\,{\rm pairs} \,,
\end{equation} 
can be obtained from the process~\eqref{eq:nf2-1f1} combined with (i.e. followed by) the same process with flavours 1 and 2 interchanged,
\begin{equation}
\label{eq:nf2-1f2}
 a_{+L}^2 \,+\, M \,\to\, \bar{b}_{\,+R}^{\,1} \,+\, M \,+\, (\bar{b}b)\,{\rm pairs} \,.
\end{equation}

\medskip

Similarly to~\eqref{eq:nf2-1f1}, \eqref{eq:nf2-2f2}, there are also anomalous processes with $n=-1$ and $n=-2$, 
which are allowed by all selection rules and the eelectrc charge conservation in this model,
\begin{eqnarray}
\label{eq:nf2-1f1-minus}
 \bar{a}_{\,-R}^1 \,+\, M &\to& b_{\,-L}^{\,2} \,+\, M \,+\, (\bar{b}b)\,{\rm pairs},\\
 \label{eq:nf2-2f2-minus}
 \bar{a}_{-R}^1 \,+\, \bar{a}_{-R}^2 \,+\,  M &\to& b_{\,-L}^{\,1} \,+\, b_{\,-L}^{\,2} \,+\,  M \,+\, (\bar{b}b)\,{\rm pairs}.
\end{eqnarray}
They are the counterparts of the corresponding processes~\eqref{eq:nf2-1f1}, \eqref{eq:nf2-2f2} with positive $n$.
One can also construct non-anomalous processes with $n=0$ by combining for example~\eqref{eq:nf2-1f2} with~\eqref{eq:nf2-1f1-minus},
\begin{equation}
\label{eq:n=0Nf2}
\bar{a}_{-R}^1 \,+\,  a_{+L}^2 \,+\, M \,\to\, \bar{b}_{\,+R}^{\,1} \,+\, b_{\,-L}^{\,2} \,+\, M \,+\, (\bar{b}b)\,{\rm pairs}.
\end{equation}
Thus we see that fermion--monopole scattering can be anomalous or non-anomalous and it can change the  $\Delta R-\Delta L$ fermion number (or $(B+L)$ in Grand Unification) by positive, negative or zero amount.
\medskip

The overall story, however,  becomes more complicated and interesting for the model with $N_f=4$ flavours which is also relevant to the SU(5) GUT theory with a single family.

\subsection{Scattering in the $\mathbf N_f=4$ model}
\label{sec:2.3}
\medskip

Let us first consider the process with two initial state fermions scattering on the monopole.  
Selecting the same initial state as in \eqref{eq:nf2-2f2} now gives,
\begin{equation}
\label{eq:nf4-2f1}
 a_{+L}^1 \,+\,a_{+L}^2 \,+\,  M \,\to\, \bar{b}_{\,+R}^{\,3} \,+\, \bar{b}_{\,+R}^{\,4} \,+\,  M \,+\, (\bar{b}b)\,{\rm pairs} \,.
\end{equation} 
The final state is unambiguously fixed by our selection rules and electric charge conservation.  
The process \eqref{eq:nf4-2f1} is consistent with the in/out selection rules of Table~\ref{Tab:1}, the anomaly selection rule \eqref{eq:anom_n} with $n=1$ and the electric charge conservation.
The expression above insures that all available fermion flavours participate in the process, as required by the anomaly (or equivalently by flavour symmetry).

In the context of the SU(5) GUT theory~\eqref{eq:nf4-2f1} is a key process for the monopole catalysis of the proton decay~\cite{Rubakov:1981rg,Callan:1982ah}, as will be reviewed in section~\ref{Sec:su5}. But what about the more elementary constituent process with a single fermion in the initial state?

Callan~\cite{Callan:1982au} was the first to study such processes in the four-flavour model and found that the final state must include half-solitons of the bosonized truncated $J=0$ theory in the monopole background. Such half-solitons (aka semitons) correspond to particles with fractional (in present context half-integer) fermion numbers. The scattering process consistent with all the selection rules in Table~\ref{Tab:1}, Eq.~\eqref{eq:anom_n} and charge conservation is~\cite{Callan:1982au}
\begin{equation}
\label{eq:nf4-f1}
 a_{+L}^1 \,+\,  M \,\to\, \frac{1}{2} \left(b_{-L}^1\,
  \bar{b}_{\,+R}^{\,2} \,
  \bar{b}_{\,+R}^{\,3} \,
  \bar{b}_{\,+R}^{\,4} \,\right) \,+\,  M \,+\, (\bar{b}b)\,{\rm pairs} \,.
\end{equation} 
The question arises of how to interpret the half-fermion particles in the final state. Since such states cannot arise 
in perturbation theory, Callan proposed that their half-integer fermion numbers should be interpreted statistically with a 50\% probability that the fermion number for a given flavour is zero ore one. The authors of~\cite{Csaki:2022qtz} argued against this and rejected the scattering process~\eqref{eq:nf4-f1} altogether based on the argument that if such massless half-fermion states existed, they would have to be true asymptotic states far from the monopole perturbation theory can be reliable applied. 
They have proposed instead the process~\cite{Csaki:2022qtz} 
\begin{equation}
\label{eq:nf4-f1Cs}
 a_{+L}^1 \,+\,  M \,\to\,
  \bar{b}_{\,+R}^{\,2} \,+\,
  \bar{b}_{\,+R}^{\,3} \,+\, 
  \bar{a}_{\,-R}^{\,4}  \,+\,  M  \,,
\end{equation} 
with a different final state that now includes three fermions and avoids half-integer fermion numbers. The final state fermion $\bar{a}_{\,-R}^{\,4}$ in this process 
cannot be in the $J=0$ state, since the selection rules in Table~\ref{Tab:1} would require it to be an incoming rather than the outgoing wave, hence the final state
of the scattering in~\eqref{eq:nf4-f1Cs} cannot be made out of three individual $J=0$ fermions. The authors of~\cite{Csaki:2022qtz}  have addressed this issue by arguing that there is a cross-entanglement between one of the final state fermions with the field angular momentum arising from one of the other fermions in the final state, so that the complete final state in~\eqref{eq:nf4-f1Cs} is a $J=0$ state. To support this statement they presented an expression for the scattering amplitude for~\eqref{eq:nf4-f1Cs} in the $J=0$ partial wave using pairwise helicities.

We disagree with the assertion of~\cite{Csaki:2022qtz} that the Callan process~\eqref{eq:nf4-f1} is invalid and that the 2D truncation in this case fails to capture the 4D physics of unitarity saturating fermion--monopole amplitudes. Our first objection is that even though the final 3-fermion state in~\eqref{eq:nf4-f1Cs} may be in the $J=0$ wave, the individual fermions are not. As such it is hard to understand how the outgoing $\bar{a}_{\,-R}^{\,4}$ fermion could be produced {\it inside} the monopole core, since it is not in a $J=0$ single particle state and would experience a very strong Coulomb repulsion from the core. We posit that the scattering process~\eqref{eq:nf4-f1}, if it exists, is suppressed by powers of $E/M_X\ll 1$ where $E$ is the energy carried by the incoming fermion.\footnote{Another way to visualise this point is to consider the inverse 
process where the three fermions $\bar{b}_{\,+R}^{\,2} \,+\,\bar{b}_{\,+R}^{\,3} \,+\, \bar{a}_{\,-R}^{\,4} $ form an initial state scattering on an anti-monopole $\overline{M}$. Since the incoming fermion $\bar{a}_{\,-R}^{\,4} $  in the $\overline{M}$ cannot be in the $J=0$ single particle state, it must bounce from the monopole core due to the centrifugal barrier.}
The point we are making does not invalidate the helicity-basis expression for the amplitude of~\eqref{eq:nf4-f1Cs} constructed in~\cite{Csaki:2022qtz}, instead it invalidates an assumption that it is unsuppressed by the monopole scale.

Our second point against rejecting the process~\eqref{eq:nf4-f1} is that it can be iterated to derive the non-controversial scattering process~\eqref{eq:nf4-2f1}
with two fermions in the initial state, in analogy to our earlier discussion~\eqref{eq:nf2-1f1}$\oplus$\eqref{eq:nf2-1f2}$\Rightarrow$\eqref{eq:nf2-2f2}
in the $N_f=2$ model.
Indeed, combining two single-fermion scattering processes~\eqref{eq:nf4-f1},
\begin{eqnarray}
\label{eq:nf4-f11}
 a_{+L}^1 \,+\,  M \,\to\, \frac{1}{2} \left(b_{-L}^1\,
  \bar{b}_{\,+R}^{\,2} \,
  \bar{b}_{\,+R}^{\,3} \,
  \bar{b}_{\,+R}^{\,4} \,\right) \,+\,  M,\\
 \label{eq:nf4-f12}
 a_{+L}^2 \,+\,  M \,\to\, \frac{1}{2} \left(b_{-L}^2\,
  \bar{b}_{\,+R}^{\,1} \,
  \bar{b}_{\,+R}^{\,3} \,
  \bar{b}_{\,+R}^{\,4} \,\right) \,+\,  M,
\end{eqnarray} 
we obtain,
\begin{eqnarray}
 a_{+L}^1 \,+\, \left(a_{+L}^2 \,+\,  M\right) &\to&
 a_{+L}^1 \,+\,  M \,+\,
 \frac{1}{2} \left(b_{-L}^2\,
  \bar{b}_{\,+R}^{\,1} \,
  \bar{b}_{\,+R}^{\,3} \,
  \bar{b}_{\,+R}^{\,4} \,\right) 
  \nonumber\\
  \label{eq:nf4-f22}
 &\to& \bar{b}_{\,+R}^{\,3} \,+\, \bar{b}_{\,+R}^{\,4}
  \,+\,  M 
  \,+\, \frac{1}{2} \left(\bar{b}_{\,+R}^{\,1} b_{-L}^1\right)
   \,+\, \frac{1}{2} \left(\bar{b}_{\,+R}^{\,2} b_{-L}^2\right), 
\end{eqnarray}
which reproduces correctly the process~\eqref{eq:nf4-2f1}.

On the other hand, if we attempt to combine two of the alternative processes of the type~\eqref{eq:nf4-f1Cs},
we would find that the desired final state on the right hand side of~\eqref{eq:nf4-2f1} is now contaminated 
by the $\bar{a}_{\,-R}^{\,3} $ and $\bar{a}_{\,-R}^{\,4} $ fermions and is not of the form~\eqref{eq:nf4-2f1} which allows only $\bar{b}$ fermions and $\bar{b}b$ pairs.
Once again this points to the fact that the corresponding combined alternative process with two fermions in the initial state contains interactions outside of the monopole core as is suppressed  by the monopole scale.

We would also like to comment on the apparent problem raised in~\cite{Csaki:2022qtz} that final states with half-integer fermion numbers, and hence the entire process~\eqref{eq:nf4-f1},
should not be allowed as such asymptotic states cannot exist in perturbation theory.
But we already know from the pairwise entanglement argument that any electrically charged states (with integer or fractional charges) cannot be considered decoupled from the monopole and cannot be described by tensor products of standard perturbative asymptotic states. 
In addition, it was also shown in~\cite{vanBeest:2023dbu} 
that the outgoing massless fermion excitations are attached to a topological surface that carries no charge or energy and which ends on the monopole. The presence of this topological surface means that the outgoing radiation does not have to have the same quantum numbers as states in the ordinary Fock space.
Similar ideas that the fermions from a `perturbatively missing' final state live in a different Fock space have also been discussed in~\cite{Kitano:2021pwt,Hamada:2022eiv}. 

In a more realistic theory, the fermions are massive. But at the distances near and inside the monopole core masses of light fermions, $m_{\rm ferm} \ll M_X$, are irrelevant and can be safely neglected, since the Coulomb terms dominate over the mass terms at small $r$. 
Only at large distances outside the monopole core the effects of light masses\footnote{And also the effects of confinement since GUT monopoles carry colour charge.}
 become relevant and 
the outgoing states with fractional fermion number become unstable and are expected to decay to states with integer fermion number~\cite{Callan:1982au,Callan:1983tm,Sen:1984qe,Preskill:1984gd,Rubakov:1988aq}.

Heavy fermion flavours with $m_{\rm ferm} \gtrsim M_X$ do not enter the monopole core and are decoupled in the bosonized formalism of Callan.

\subsection{Scattering amplitudes with pairwise helicities}
\label{sec:2.4}
\medskip

To provide some concrete evidence in favour of the scattering processes of the type~\eqref{eq:nf4-f1} involving fractional fermions
here we will construct their on-shell amplitude expressions using a combination of single particle and pairwise helicities.

\medskip

When both electric $e_i$ and magnetic $g_{Mi}$ charges are present the asymptotic state of the $S$ matrix are not given by tensor products of single-particle states.
More than 50 years ago  Zwanziger~\cite{Zwanziger:1972sx} pointed out that each pair of particles $(i,j)$ with non-vanishing $q_{ij}=e_i g_{Mj} - e_j g_{Mi} \neq 0$ constitutes an entangled pairwise state that carries a non-vanishing angular momentum of the electromagnetic field for the pair $(i,j)$. 
The DSZ quantization rule~\cite{Dirac:1931kp,Schwinger:1975ww,Zwanziger:1968rs}, which follows from the quantization of this angular momentum, implies that $q_{ij}$ is quantized in half integer units. 
In particular, for a fermion--monopole pair we have $q_{fM} = \pm 1/2$ in agreement with Eq.~\eqref{eq:qdef}.

The S matrix formalism for electric--magnetic scattering pioneered by
Cs\'aki~{\it etal} in~\cite{Csaki:2020inw} 
identified $q_{ij}$ as the pairwise helicity -- a half-integer variable that characterises each entangled $(i,j)$ pair, and introduced the corresponding 
pairwise helicity spinors $|p^{\flat \pm}_{ij}\rangle$ and  $|p^{\flat \pm}_{ij}]$.
It then follows that that under Lorentz transformations asymptotic multi-particle states pick up an extra little group phase factor $e^{i q_{ij} \phi_{ij}}$
for each electric--magnetic pair. For example, a two-particle out-state with with an electrically charged particle $i$ and a monopole $M$ with momenta 
$p_i$, $p_M$ and spins $s_i$, $s_M$ 
transforms as~\cite{Csaki:2020inw},
\begin{equation}
U(\Lambda) |p_i,p_M; s_i, s_M; q_{iM}\rangle\,=\, e^{i q_{iM} \phi_{iM}} \,  |\Lambda p_i,\Lambda p_M; s'_i, s'_M; q_{iM}\rangle \, {\cal D}_{s'_i s_i} {\cal D}_{s'_M s_M}\,,
\end{equation}
where $U(\Lambda)$ is the unitary representation of the Lorentz transformation $\Lambda$, the phase factor $e^{i q_{iM}\phi_{iM}}$ is the action of the little group 
associated with the momentum pair $(p_i, p_M)$, and ${\cal D}_{s'_i s_i}$,  ${\cal D}_{s'_M s_M}$ are the individual single particle little group factors. 
For an in-state the pairwise little group phases are opposite, so that the amplitude
satisfies~\cite{Csaki:2020inw},
\begin{eqnarray}
\label{eq:L_ampl}
&&\tilde{\cal A} (p_1, \ldots, p_n, p_M |\, k_1, \ldots, k_m, k_M) = 
\\
&&\qquad \qquad e^{i \sum_{i=1}^n q_{iM} \phi_{iM}} \, e^{i \sum_{l=1}^m q_{lM} \phi_{lM}}
{\cal A} (\Lambda p_1, \ldots, \Lambda p_n, \Lambda p_M |\, \Lambda k_1, \ldots, \Lambda k_m, \Lambda k_M)
\,. \nonumber
\end{eqnarray}
The amplitude above is written for $n$ particles scattering into $m$ particles on a {\it single} monopole and we have also absorbed the 
 single-particle little group ${\cal D}$-factors into the expression $\tilde{\cal A}$ on the left hand side.

To construct the amplitude $\tilde{A}$ which transforms according to~\eqref{eq:L_ampl}, we make use of the pairwise helicity spinors constructed in~\cite{Csaki:2020inw}
(which for completeness we briefly review in the Appendix) and their Lorentz transformation properties,
\begin{equation}
\label{eq:pb>tr}
\Lambda \, |p^{\flat \pm}_{ij}\rangle \,=\, e^{ \pm \frac{i}{2}\phi(p_i,p_j,\Lambda)} \, |\Lambda p^{\flat \pm}_{ij}\rangle 
\end{equation}
\begin{equation}
\label{eq:pb]tr}
[p^{\flat \pm}_{ij} | \, \tilde\Lambda \,=\, e^{ \mp \frac{i}{2}\phi(p_i,p_j,\Lambda)} \, [\Lambda p^{\flat \pm}_{ij} | 
\end{equation}
where $\Lambda$ and $\tilde\Lambda$ on the left hand sides represent Lorentz transformations in spinor-bases,  and on the right --  in momentum-basis.

\medskip

We now proceed to construct an expression for the amplitude $\tilde{\cal A}$ for the minimal process~\eqref{eq:nf4-f11} where a single fermion is scattering on a scalar monopole in the $J=0$ partial wave.
Using the standard all-outgoing conventions for amplitude momenta, the contribution to the amplitude $\tilde{\cal A}$ from the incoming state $a_{+L}^1 + M $ is given by,
\begin{equation}
\label{eq:in}
 (a_{+L}^1+  M)_{\rm in} \,\, \Rightarrow \qquad   [a_{+L}^1 | \, p^{\flat -}_{a^1 M} ]\,,
\end{equation}
where the single-particle helicity spinor
$|a_{+L}^1 ]$  is the standard wave-function representing an incoming $a_{+L}^1$ fermion in the all-outgoing momentum convention.\footnote{An outgoing particle 
with momentum $p_i^\mu$ and helicity $h$ contributes to an amplitude a factor of $|i\rangle^{-2h}$ or equivalently $|i]^{2h}$, see 
e.g.~\cite{Dixon:1996wi,Arkani-Hamed:2017jhn}. For example, a negative-helicity $h=-1$ outgoing gluon contributes to the amplitude an overall factor of  $|i\rangle^{2}$ and a positive-helicity gluon -- a factor of $|i]^{2}$, in agreement with the well-known Parke-Taylor MHV amplitudes. An outgoing left-handed fermion has $h=-1/2$ and gives $|i\rangle^{1}$, while  
an incoming left-handed fermion in the all-outgoing convention counts for $h={+1/2}$ and contributes $1/ |i\rangle \,\sim | i]$. }
The second factor,
$| p^{\flat -}_{a^1 M} ]$, is the pairwise helicity spinor associated with the $(a^1,M)$ pair.

The expression on the right hand side of~\eqref{eq:in} is uniquely determined by the requirements
that: its helicity spinors can involve only the initial states; all (Lorentz) spinor indices must be contracted as this is a $J=0$ state; Lorentz transformations of the pairwise
helicity spinor $| p^{\flat -}_{a^1 M} ]$ should give the phase factor $e^{\, i q_{a^1M} \, \phi}$, to be consistent with~\eqref{eq:L_ampl}, which is indeed the case thanks 
to~\eqref{eq:pb>tr} and the fact that $q_{a^1M}=1/2$ according to Table~\ref{Tab:1}.  The expression~\eqref{eq:in} for the in-state contribution of a  
single-fermion--single-monopole pair is in agreement with the result in~\cite{Csaki:2022qtz}. 

We next should determine the contribution to of the outgoing state to the amplitude $\tilde{\cal A}$.
For each half-fermion in the monopole background we take,
\begin{eqnarray}
\label{eq:out1}
 (\frac{1}{2} b_{-L}^1 +  M)_{\rm out} \,\, \Rightarrow \qquad   && \sqrt{\langle b_{-L}^1 | \,p^{\flat -}_{b^1 M} \rangle} \,,\\
 \label{eq:outi}
 (\frac{1}{2} \bar{b}_{+R}^{\,i} +  M)_{\rm out} \,\, \Rightarrow \qquad   &&\sqrt{[\bar{b}_{+R}^{\,i} | \, p^{\flat -}_{\bar{b}^i M} ]}\,, \quad i=2,3,4.
\end{eqnarray}
It is easy to verify each of these factors transforms with the correct pairwise little group phase $e^{\, i q \, \phi}$, as required by~\eqref{eq:L_ampl},
\begin{eqnarray}
\nonumber
  \sqrt{\langle b_{-L}^1 | \,p^{\flat -}_{b^1 M} \rangle} \,\to\, e^{\,-\frac{i}{4} \phi(b^1,M)}\,
  \sqrt{\langle \Lambda b_{-L}^1 | \Lambda p^{\flat -}_{b^1 M} \rangle} &=&
  e^{\,{i} q(\frac{1}{2}b^1,M) \,\phi(b^1,M)}\,
  \sqrt{\langle \Lambda b_{-L}^1 | \Lambda p^{\flat -}_{b^1 M} \rangle}
  \,,\\
  \nonumber
  \sqrt{[\bar{b}_{+R}^{\,i} | \, p^{\flat -}_{\bar{b}^i M} ]} \,\to\, e^{\frac{i}{4} \phi(b^1,M)}\,
\sqrt{[ \Lambda \bar{b}_{+R}^{\,i} | \Lambda  p^{\flat -}_{\bar{b}^i M} ]} &=&
  e^{\,{i} q(\frac{1}{2}\bar{b}^i,M) \,\phi(\bar{b}^i,M)}\,
  \sqrt{[ \Lambda \bar{b}_{+R}^{\,i} | \Lambda  p^{\flat -}_{\bar{b}^i M} ]}\,,
\end{eqnarray}
where we used $q=-\frac{1}{4}$ for the $(\frac{1}{2} b^1,M)$ pair, and $q=\frac{1}{4}$ for $(\frac{1}{2}\bar{b}^{\,i},M)$ by adapting
 the $q$-charges in Table~\ref{Tab:1} to half-fermions which have half of the electric charges of the corresponding fermions. 

Our final result for the amplitude of the scattering process~\eqref{eq:nf4-f11} on a scalar monopole in the $J=0$ wave is,
\begin{equation}
\label{eq:nf4-f1Amp}
 \tilde{\cal A}_{\, \rm Eq.\eqref{eq:nf4-f11}} \propto
  [a_{+L}^1 | \, p^{\flat -}_{a^1 M} ]\left(\langle b_{-L}^1 | \,p^{\flat -}_{b^1 M} \rangle\,
  [\bar{b}_{+R}^{\,2} | \, p^{\flat -}_{\bar{b}^2 M} ]\,
   [\bar{b}_{+R}^{\,3} | \, p^{\flat -}_{\bar{b}^3 M} ]\,
    [\bar{b}_{+R}^{\,4} | \, p^{\flat -}_{\bar{b}^4 M} ]
  \right)^{1/2}.
\end{equation} 
The above expression for the amplitude can also be re-written in a dimensionless form corresponding to the case were all incoming and outgoing states were 
normalised to 1; 
in this case each of the spinor products for the would be divided by the energy $E_i=|{\bf p}_{i}|$ of the corresponding electrically charged particle.
We also note that all pairwise helicity spinors in~\eqref{eq:nf4-f1Amp} are of the $p^{\flat -}_{ij}$ type, hence when combined with the $i$-particle spinor, 
their spinor products avoid the trivially vanishing combinations~\eqref{eq:ang0}-\eqref{eq:sq0}.

\medskip

Expressions for amplitudes with semiton states and the fact that they involve square roots of spinor products (and more generally $2/N_f$ roots for the model with $N_f$ flavours) is one of our main results. An appearance of such fractional $2/N_f$ powers is in fact in agreement with the interpretation of the fermion--monopole scattering 
proposed in~\cite{vanBeest:2023dbu}.
In their picture the outgoing state with fractional fermion number is composed from a single fermion and a twist operator that sits at the end of a topological surface connecting the outgoing state to the monopole. On the two-dimensional $(r,t)$-plane the topological surface is a string that manifests itself as a brunch cut. This picture is confirmed by the appearance the same brunch cut resulting from the fractional powers of spinor products in our expressions for scattering amplitudes involving a semiton.

\medskip

The amplitude for the companion process~\eqref{eq:nf4-f12} is obtained from~\eqref{eq:nf4-f1Amp} by interchanging the flavour labels $1 \leftrightarrow 2$,
\begin{equation}
\label{eq:nf4-f122Amp}
 \tilde{\cal A}_{\, \rm Eq.\eqref{eq:nf4-f12}} \propto
  [a_{+L}^2 | \, p^{\flat -}_{a^2 M} ]\left(\langle b_{-L}^2 | \,p^{\flat -}_{b^2 M} \rangle\,
  [\bar{b}_{+R}^{\,1} | \, p^{\flat -}_{\bar{b}^1 M} ]\,
   [\bar{b}_{+R}^{\,3} | \, p^{\flat -}_{\bar{b}^3 M} ]\,
    [\bar{b}_{+R}^{\,4} | \, p^{\flat -}_{\bar{b}^4 M} ]
  \right)^{1/2}.
\end{equation} 
The combination of the processes~\eqref{eq:nf4-f11} and \eqref{eq:nf4-f12} should give the scattering process~\eqref{eq:nf4-f22} 
with two fermions in the initial state.
Taking the product of the amplitudes~\eqref{eq:nf4-f1Amp} and~\eqref{eq:nf4-f122Amp}, we find
\begin{eqnarray}
 \tilde{\cal A}_{\, \rm Eq.\eqref{eq:nf4-f22}} \propto &&
 [a_{+L}^1 | \, p^{\flat -}_{a^1 M} ]\,  [a_{+L}^2 | \, p^{\flat -}_{a^2 M} ]\,
 [\bar{b}_{+R}^{\,3} | \, p^{\flat -}_{\bar{b}^3 M} ]\,
    [\bar{b}_{+R}^{\,4} | \, p^{\flat -}_{\bar{b}^4 M} ] \times \nonumber\\
\label{eq:Amp2fbbar}    
&& \,\,\left(
 \langle b_{-L}^1 | \,p^{\flat -}_{b^1 M} \rangle\,
  [\bar{b}_{+R}^{\,2} | \, p^{\flat -}_{\bar{b}^2 M} ]\,
  \langle b_{-L}^2 | \,p^{\flat -}_{b^2 M} \rangle\,
  [\bar{b}_{+R}^{\,1} | \, p^{\flat -}_{\bar{b}^1 M} ]\,
     \right)^{1/2}.
\end{eqnarray} 

It is worthwhile to point out that when we combine two square roots, for example one factor of $([\bar{b}_{+R}^{\,3} | \, p^{\flat -}_{\bar{b}^3 M} ])^{1/2}$
from the amplitude~\eqref{eq:nf4-f1Amp} and another factor of $([\bar{b}_{+R}^{\,3} | \, p^{\flat -}_{\bar{b}^3 M} ])^{1/2}$ from the second amplitude~\eqref{eq:nf4-f122Amp},
they each correspond to the same semiton but with potentially different on-shell momenta, $p_1$ and $p_2$, as they occur in two independent processes. But since we 
are working with the $J=0$ spherical waves, the four-momenta of the two semitons are actually proportional to each other. This implies that the spinor helicities
 $([\bar{b}_{+R}^{\,3} (p_1) |$ and $([\bar{b}_{+R}^{\,3} (p_2) |$ are also the same up to a proportionality coefficient. It then follows that for an appropriately normalised 
 amplitude we can combine the two square roots into a single expression $[\bar{b}_{+R}^{\,3} | \, p^{\flat -}_{\bar{b}^3 M} ]$, which is what we have done in deriving 
  the result for the amplitude~\eqref{eq:nf4-f122Amp}.

We also note that processes with no $\bar{b}b$ pairs in the out state are given by omitting the second line in~\eqref{eq:Amp2fbbar}, thus we have
\begin{equation}
\label{eq:prAmp2f}
 a_{+L}^1 \,+\,a_{+L}^2 \,+\,  M \,\to\, \bar{b}_{\,+R}^{\,3} \,+\, \bar{b}_{\,+R}^{\,4} \,+\,  M,
 \end{equation}
\begin{equation}
\label{eq:Amp2f}
\qquad \qquad
\tilde{\cal A}
 \propto
 [a_{+L}^1 | \, p^{\flat -}_{b^1 M} ]\,  [a_{+L}^2 | \, p^{\flat -}_{b^2 M} ]\,
 [\bar{b}_{+R}^{\,3} | \, p^{\flat -}_{\bar{b}^3 M} ]\,
    [\bar{b}_{+R}^{\,4} | \, p^{\flat -}_{\bar{b}^4 M} ].
 \end{equation} 

\medskip

Finally, following analogous steps to the approach described above, we can construct amplitudes for the processes 
with negative or vanishing (i.e. non-anomalous) $n$, such as:
\begin{eqnarray}
\label{eq:f1-minus}
&& \bar{a}_{\,-R}^4 \,+\, M \,\to\, \frac{1}{2} \left(
 \,\bar{b}_{\,+R}^{\,\,4} \,
  b_{-L}^1\,b_{-L}^2\,b_{-L}^3
   \,\right) \,+\,  M,\\
\label{eq:f2-minus}
&& \bar{a}_{\,-R}^1 \,+\,\bar{a}_{\,-R}^2 \,+\, M \,\to\,   b_{-L}^3\,+\,b_{-L}^4\,+\, M,\\
 \label{eq:f2-zero}
&& \bar{a}_{-R}^1 \,+\,  a_{+L}^2 \,+\, M \,\to\,  \bar{b}_{\,+R}^{\,1} \,+\, b_{\,-L}^{\,2} \,+\, M,
\end{eqnarray} 
that we will need in the following section.

\medskip

Analytic expressions we have derived for the amplitudes above are presented with a proportionality signs. The so-far missing kinematic factors on the right hand side these amplitudes can nevertheless be fixed on dimensional grounds and from the requirement that the lowest partial wave anomalous amplitudes for scattering processes saturate unitarity bounds.

\section{Scattering of fermions with SU(5) GUT monopoles}
\label{Sec:su5}

In the minimal GUT theory the 't Hooft--Polyakov monopole lives in the $SU(2)_M$ subgroup of the $SU(5)_{\rm GUT}$.
We consider a single generation of massless fermions in this model.
Left-handed Weyl fermions transform in the $\bar{5}$ and $10$ representations of $SU(5)_{\rm GUT}$ are represented by $N_f=4$ of $SU(2)_M$ 
doublets~\cite{Rubakov:1988aq},
\begin{equation}
\label{eq:fLsu5}
\psi_L^{\,i}\,=\, \begin{pmatrix}
a_{+}^{\,i} \\
b_{-}^{\,i}
\end{pmatrix}_L\,\Rightarrow\,
\begin{pmatrix}
\bar{u}^{1}_{L} \\
u^2_{L}
\end{pmatrix},\,
\begin{pmatrix}
-\bar{u}^{\,2}_{\,L} \\
u^{\,1}_{\,L}
\end{pmatrix},\,
\begin{pmatrix}
d^{3}_{L} \\
\,\bar{e}_{\,L}
\end{pmatrix},\,
\begin{pmatrix}
e_{L} \\
-\bar{d}^{\,3}_{\,L}
\end{pmatrix},
\end{equation}
with the corresponding anti-fermions given by,\footnote{To be clear, $\bar{u}^{1}_{L}$ in our notation describes the Dirac conjugate of an idependent Right-handed field, $\bar{u}^{1}_{L}=\overline{({u}^{1}_{R}})$. Thus the components $u_L^1$ and $\bar{u}^{1}_{L}$ in \eqref{eq:fLsu5} are independent fields; their anti-particle counterparts are $u_R^1$ and $\bar{u}^{1}_{R}$  listed in \eqref{eq:fRsu5} }
\begin{equation}
\label{eq:fRsu5}
\bar\psi_{\,\,R}^{\,i}\,=\, \begin{pmatrix}
\bar{b}_{+}^{\,i}  \\
\bar{a}_{-}^{\,i}
\end{pmatrix}_R
\,\Rightarrow\,
 \begin{pmatrix}
\bar{u}^{\,2}_{\,R} \\
u^{1}_{R}
\end{pmatrix},\, 
 \begin{pmatrix}
\bar{u}^{\,1}_{\,R} \\
-u^{2}_{R}
\end{pmatrix},\,
\begin{pmatrix}
e_{R} \\
\bar{d}^{\,3}_{\,R}
\end{pmatrix},\,
\begin{pmatrix}
-d^3_{R} \\
\bar{e}_{\,R}
\end{pmatrix},
\end{equation}
and all other fermions are  $SU(2)_M$-singlets.
The electric charges $e_M$ entering the DSZ quantization rule are not the actual QED electric charges of the up and down quarks and electrons, but the 
charges of $T^3$ of the $SU(2)_M$ subgroup of the $SU(5)_{GUT}$ where the 't Hooft-Polyakov monopole is embedded. They coincide with the electric charges of $a_+$ and $b_-$ fermions in the SU(2) Georgi-Glashow model considered in the previous section. These are the electric charges under unbroken $U(1)_M \in SU(2)_M$ following the electroweak symmetry breaking. The magnetic charge of the monopole under the $U(1)_M$ is $g_M=1$ as before. Note, however that $U(1)_{QED}$ electric charges of the fermions all have opposite signs relative to their $e_M$ charges, and the same change of sign also applies to the $U(1)_{QED}$ magnetic charge of the GUT monopole, which becomes $g^{mg}_{QED} =-1$. 

\medskip

In analogy with Table~\ref{Tab:1} we summarise the $U(1)_M$ charges and the in/out properties of the $J=0$ wave fermion components in the two Tables below.

 \begin{table}[ht]
\centering
\begin{tabular}{c c c c c c c}
\hline\hline
 & $\psi_L^1$ & $\psi_L^2$ &$\psi_L^3$ &$\psi_L^4$ &$\,\,q=e_M g_M$ & \,\,in/out  \\ [0.5ex] 
\hline
$a_{\,+L}:$  & $\bar{u}^{1}_{L}$  &$\bar{u}^{2}_{L}$  &$d^{3}_{L} $ & $e_{L}$  & $1/2$ &{in} \\
$b_{\,-L}:$  & $u^{2}_{L}$  &$u^{1}_{L}$  &$\bar{e}_{\,L}$ & $\bar{d}^{\,3}_{\,L}$  &  $-1/2$ & {out} \\
[1ex]   
\hline\\
\end{tabular}
\caption{GUT left-handed fermions~\eqref{eq:fLsu5}, their $U(1)_M$ pairwise helicity charges $q$ and the in/out states assignment in the $J=0$ wave.}
\label{Tab:2}
\end{table}

\begin{table}[ht]
\centering
\begin{tabular}{c c c c c c c}
\hline\hline 
& $(\overline{\psi})_{R}^1$ &  $(\overline{\psi})_{R}^2$ & $(\overline{\psi})_{R}^3$ & $(\overline{\psi})_{R}^4$ &$\,\,q=e_M g_M$ & \,\,in/out \\ [0.5ex] 
\hline
 $\bar{b}_{\,+R}:$ & $\bar{u}^{\,2}_{\,R}$  &$\bar{u}^{\,1}_{\,R}$  &$e_{R}$ & $d^3_{R} $  & $1/2$ &out \\
$\bar{a}_{\,-R}:$  & $u^{1}_{R}$  &$u^{2}_{R}$  &$\bar{d}^{\,3}_{\,R}$ & $\bar{e}_{\,R}$  & $-1/2$ &in \\
[1ex]   
\end{tabular}
\caption{GUT anti-fermions~\eqref{eq:fRsu5}}
\label{Tab:3}
\end{table}

Scattering processes~\eqref{eq:prAmp2f}  and~\eqref{eq:f2-minus} with two fermions in the initial state, along with their amplitudes in the $J=0$ wave are given by
\begin{eqnarray}
\label{eq:3.5}
 &&\bar{u}^{1}_{L} \,+\,\bar{u}^{2}_{L}\,+\,  M \,\to\, d^3_{R} \,+\, {e}_{R} \,+\,  M,
\\
\label{eq:3.6}
 &&\tilde{\cal A}
 \propto
 [\bar{u}_{\,L}^{1} | \, p^{\flat -}_{\bar{u}^{1} M} ]\,  [\bar{u}_{\,L}^{2}| \, p^{\flat -}_{\bar{u}^{2} M} ]\,
 [d_R^{3} | \, p^{\flat -}_{d^3 M} ]\,
    [e_R | \, p^{\flat -}_{{e} M} ],
    \end{eqnarray} 
and 
\begin{eqnarray}
\label{eq:3.3}
&& u^{1}_{R} \,+\, u^{2}_{R} \,+\, M \,\to\,   
\bar{d}^{\,3}_{\,L}\,+\,\bar{e}_{\,L}\,+\, M,\\
\label{eq:3.4}
 &&\tilde{\cal A} 
 \propto
 \langle u_R^{1} | \, p^{\flat -}_{u^{1} M} \rangle\,  \langle u_R^{2} | \, p^{\flat -}_{u^{2} M} \rangle\, 
 \langle \bar{d}_{\,L}^{3} | \, p^{\flat -}_{\bar{d}^3 M} \rangle\,
    \langle\bar{e}_{\,L} | \, p^{\flat -}_{\bar{e} M} \rangle,
\end{eqnarray} 
which describe the monopole catalysis of anti-proton and proton decays respectively.

Scattering processes with a single fermion in the initial state of the type~\eqref{eq:nf4-f11} and~\eqref{eq:f1-minus} in the GUT theory result in 
\begin{eqnarray}
\label{eq:3.7}
 &&\bar{u}_{\,L}^{\,1} \,+\,  M \,\to\, \frac{1}{2} \left(u_{L}^2\,
  \bar{u}_{\,R}^{\,1} \,
 e_{R} \,
 d_{R}^{\,3} \,\right) \,+\,  M,\\
 \label{3.8}
 &&\tilde{\cal A}
 \, \propto\, 
  [\bar{u}_{\,L}^{\,1}  | \,p^{\flat -}_{{u}^1 M} ]
  \left(
  \langle {u}_{\,L}^{\,2} | \,p^{\flat -}_{{u}^2 M} \rangle\,
  [\bar{u}_{R}^{\,1} | \, p^{\flat -}_{\bar{u}^1 M} ]\,
   [e_{R} | \, p^{\flat -}_{e M} ]\,
    [d_{R}^{3} | \, p^{\flat -}_{d^3 M} ]
  \right)^{1/2}
\end{eqnarray} 
and 
\begin{eqnarray}
\label{eq:3.9}
&& \bar{e}_{\,R} \,+\, M \,\to\, \frac{1}{2} \left(
 \,d_{R}^{\,3} \,
 u_{L}^2\,u_{L}^1\,\bar{e}_{L}
   \,\right) \,+\,  M,\\
\label{eq:3.10}
&&\tilde{\cal A}
 \, \propto\, 
  \langle \bar{e}_{R} | \, p^{\flat -}_{\bar{e} M} \rangle
  \left(
  [ d_{R}^{\,3} | \,p^{\flat -}_{d^3 M} ]\,
  \langle u^2_L | \, p^{\flat -}_{u^2 M} \rangle\,
   \langle u^1_L | \, p^{\flat -}_{u^1 M} \rangle\,
    \langle\bar{e}_{L} | \, p^{\flat -}_{\bar{e} M} \rangle
  \right)^{1/2}
\end{eqnarray} 
where the process~\eqref{eq:3.9} and its amplitude~\eqref{eq:3.10} describe the anomalous positron--monopole reaction studied originally by Callan 
with a production of 
half-fermions with the quantum numbers of a half-proton and half-positron.

And finally, for the non-anomalous process~\eqref{eq:f2-zero} in our $SU(5)_{\rm GUT}$ 1-family model we have
\begin{eqnarray}
 \label{eq:3.11}
&& u_{R}^1 \,+\,  \bar{u}_{\,L}^{\,2} \,+\, M \,\to\,  \bar{u}_{\,R}^{\,2} \,+\, u_{L}^{1} \,+\, M,\\
\label{eq:3.12}
&&\tilde{\cal A}\, \propto\, 
 \langle u_R^{1} | \, p^{\flat -}_{u^{1} M} \rangle\,  [\bar{u}_{\,L}^{2}| \, p^{\flat -}_{\bar{u}^{2} M} ]\,
 [\bar{u}_{\,R}^{2}| \, p^{\flat -}_{\bar{u}^{2} M} ]\,\langle u_L^{1} | \, p^{\flat -}_{u^{1} M} \rangle.
\end{eqnarray} 

\section{Conclusion}
\label{Sec:concl}

In this paper we have re-examined the effects and properties of scattering processes involving massless fermions and magnetic monopoles. 
We described these processes and constructed the corresponding on-shell amplitudes first in the minimal $SU(2)$ model that supports 't Hooft--Polyakov monopoles and also in the $SU(5)$ GUT theory with a single family of massless fermions.
These processes are unsuppressed, they do not depend on the monopole or the $SU(5)_{\rm GUT}$ mass scales scale even at low energies; they are instrumental for the monopole catalysis of proton decay and are also interesting on their own right. We derived helicity amplitudes 
for fermion--monopole scattering in events with a single fermion in the initial state and fractional fermion numbers in the final state, and provided non-trivial tests on such processes by combining them to reproduce the amplitudes for processes with more fermions in the initial- and integer fermion numbers in the final state. 
A variety of different fermion--monopole scattering processes and  scattering amplitudes were presented including those with and without the
anomalous $(B+L)$-violation.
Kinematic-dependent prefactors for the amplitude expressions derived in this paper can also be determined  from the requirement that the lowest partial wave anomalous amplitudes saturate unitarity bounds, and will be presented elsewhere~\cite{WIP}.

\section*{Acknowledgements}

This work is supported by the STFC under consolidated grant ST/P001246/1.

\newpage 
\appendix

\section{Appendix: Pairwise spinors}
\label{app:Son}

\noindent For every pair of particles with the momenta $p_i$ and $p_j$ their pairwise spinors are defined as follows~\cite{Csaki:2020inw}.
First the momentum pair is Lorentz boosted into the CoM frame,
\begin{equation}
k_i = (E_i,0,0,p_c)\,, \quad k_j = (E_j,0,0,-p_c)\,
\end{equation}
where the positive and negative pairwise null-momentum vectors are defined,
\begin{equation}
\label{eq:defkflat}
k^{\flat \pm}_{ij}\,=\, p_c\, (1,0,0,\pm 1).
\end{equation}
This is used to introduce the pairwise helicity spinors in the CoM frame,
\begin{equation}
k^{\flat \pm\, \mu}_{ij}\, \sigma_{\mu\, \alpha, \dot{\alpha}}\,=\, {|k^{\flat \pm}_{ij}\rangle}_{\alpha}\,{ [k^{\flat \pm}_{ij}|}_{\dot\alpha}.
\end{equation}
These can now be boosted back to a general Lorentz frame with the spinorial Lorentz transformations~\cite{Csaki:2020inw},
\begin{equation}
 {|p^{\flat \pm}_{ij}\rangle}_{\alpha}\,=\,\Lambda_{\alpha}^{\,\, \beta}\,{|k^{\flat \pm}_{ij}\rangle}_{\beta}
 \,\,, \qquad
  { [p^{\flat \pm}_{ij}|}_{\dot\alpha}\,=\, 
 { [k^{\flat \pm}_{ij}|}_{\dot\beta} \, \tilde\Lambda^{\dot\beta}_{\,\, \dot\alpha}.
\end{equation}
The resulting pairwise helicity spinors $[p^{\flat \pm}_{ij}|$ and $[p^{\flat \pm}_{ij}|$ satisfy the Lorentz transformation 
properties in~\eqref{eq:pb>tr}-\eqref{eq:pb]tr}.

The defining equation \eqref{eq:defkflat} for the null pairwise momenta in the CoM frame implies 
that $k^{\flat +}_{ij}$ is collinear with the null momentum of the $i$ particle $k_i$, and $k^{\flat -}_{ij}$ is aligned with $k_j$. 
This results in the following vanishing identities for the single-particle and pairwise helicities~\cite{Csaki:2020inw},
\begin{eqnarray}
\label{eq:ang0}
&\langle i\, p^{\flat +}_{ij}\rangle\,=\,0 \,, 
 \quad &\langle j\, p^{\flat -}_{ij}\rangle \,=\,0 \,,\\
 \label{eq:sq0}
 &[ i\, p^{\flat +}_{ij}]\,=\,0 \,, 
 \quad &[ j\, p^{\flat -}_{ij}] \,=\,0.
\end{eqnarray}

\bigskip

\bibliographystyle{inspire}
\bibliography{main}

\clearpage
\appendix
\include{variables}

\end{document}